# Designing Inverse Dynamic Controller with Integral Action for Motion Planning of Surgical Robot in the Presence of Measurable Disturbances

A. Aminzadeh Ghavifekr[*], M.A. Badamchizadeh[**], G. Alizadeh[***], A. Arjmandi[****]

[*] Department of Electrical and Computer Engineering, University of Tabriz, Tabriz, Iran
Email: a_aminzadeh89@ms.tabrizu.ac.ir
[**] Department of Electrical and Computer Engineering, University of Tabriz, Tabriz, Iran
Email: mbadamchi@tabrizu.ac.ir
[***] Department of Electrical and Computer Engineering, University of Tabriz, Tabriz, Iran
Email: alizadeh@tabrizu.ac.ir
[****] Department of Electrical and Computer Engineering, K.N. Toosi University, Tehran, Iran
Email: a.arjmandi@ee.kntu.ac.ir

*Abstract:* *Robotic laparoscopic grasper is a surgical tool with minimal invasion. In this robot, achieve goals like precise tracking, stability and disturbance rejection are very important. In this paper, first the stages of modeling and simulating of laparoscopic robot will be discussed and the reasons for selecting the appropriate materials for different parts of proposed practical robot will be explained. It would be required this robot, which will do the main part of the surgery, be controlled based on the uncertain properties of the tissues of patients body.*
*Inverse dynamic controller with integral action is applied to improve the accuracy of tracking procedure for a surgical manipulator to track a specified reference signal in the presence of tremor that is modeled as constant bounded disturbance. Based on the disturbance rejection scheme, tracking controller is constructed which is asymptotically stabilizing in the sense of Lyapunov. It is shown that how under proper assumptions; the selected schemes succeed in achieving disturbance rejection at the input of a nonlinear system. Computer simulation results demonstrate that accurate trajectory tracking can be achieved by using the proposed controller.*

## 1. Introduction

There have been numerous studies in literature in the field of surgical robots [1, 2]. Indeed in these systems the surgeon's skill and intelligence are combined with the robot's precision, repeatability and its reach ability to the sensitive points those are not within reach of the surgeon.
One of methods that is used in this field is laparoscopic operation [3]. In this method using a small cavity which is made on the body of patient, camera and surgery instruments are sent into body and special monitoring is used to control and evaluate its trajectory. Due to the small cavity created on the body, recovery period becomes short, but due to the lack of a physical interaction between surgeon and tissues of patients' body, robotic structures have attracted much attention.
These robots not only increase the accuracy and simplicity of surgery, but also avoid unwanted hurts and injuries to the patients' body by using haptic structure and taking force feedback [4],[5]. One of the important parts of surgical robots is their grasper which is in contact with the tissues of the patients' body. Making autonomous graspers of surgical robots would play a major role in reducing the errors of the surgical operations and would improve the surgeons' perception of the region of operation [6].
One of the primary studies in the field of dynamic modeling of surgical robots was done by Chirikjian [7]. In this study, using geometric properties of three dimensional curves and utilizing Extended Frenet-Serret frames, an approximated method is proposed for parameterization of corresponding curve of central skeleton.
The problem of rejecting disturbances occurring in dynamical systems is a fundamental problem in control theory, with numerous technological applications such as control of vibrating structures [8], active noise control [9] and control of rotating mechanisms [10].
Since many objectives in control engineering practice involve signal peak and the disturbance signals of the plants are constant bounded in most cases, many papers have dealt with the problem of constant bounded disturbance rejection without delay. In [11] this problem has been discussed for nonlinear systems.
The tracking control of robotic manipulators has been extensively studied. Trajectory tracking errors for robotic system are subjected to various disturbances, such as measurement and modeling error and load variance [12]. In order to obtain better tracking performance the developed control algorithms should be capable of reducing these uncertainties effects.

If the parameters of the robot are completely known, feedback linearization technique or computed torque scheme can be used for control design. When system parameters are unknown and vary in wide ranges, adaptive and robust control algorithms need to be applied [13].

There are several inherent difficulties associated with these approaches. First of all, these designs require knowledge of the structure of the robot, which may not be available. It has also be demonstrated [l4-15] that these adaptive controllers may have lack of robustness against unmodeled dynamics, sensor noise, and other disturbances.

Tremor of surgeons' hands and hysteresis phenomena are two major problems in controlling the end-effectors of surgical robots that lead to considerable errors. To compensate hand tremor, there are two approaches: tremor model based and robust approach. In the tremor model based approach, it is required to precisely model tremor of hand to calculate amount of this resistive force to be added to the applied force of the controller, so the effect of tremor can be reduced. The major drawback of this method is the necessity for precisely modeling of the tremor. Actual proposed models are too complex to be applied when designing a controller. In this paper, the second method will be used to deal with the problem of eliminating the effect of hand tremors of surgeon.

The effect of these tremors in dynamic equations can be represented by a term summed with systems input, which is called disturbance in control literature. So applying robust control methods, it is possible to eliminate the terms added to input. In this paper Inverse-dynamic controller with an integral action is used to eliminate the additive terms and to cancel uncertain constant-bounded disturbances like load torque. Considering tremor of surgeon's hand properties, using this method is favorable due to the unknown but bounded nature of tremors.

The rest of this paper is organized as follows: In section 2 physical and practical structure of laparoscopic grasper is explained. Section 3 describes dynamic model and geometric calculation of grasper. In section 4 the robotic tracking problem is formulated into a proper tremor rejection problem and Inverse dynamic control with an integral action method is presented to reject the tremor of surgeons' hand that is considered as constant bounded disturbance. All system uncertainties are lumped into the disturbance term. In order to prove the effectiveness of the proposed controllers, simulation results have been presented in section 5.

## 2. Physical structure of laparoscopic grasper

In designing part of the laparoscopic robot, first an initial model of the robot is developed and the required materials and tools for designing the robot are selected. Figure. 1, depicts the simulation model of designed robotic laparoscopic grasper.

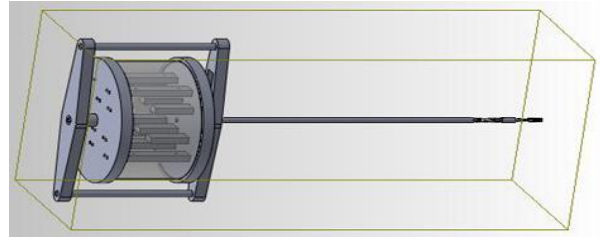

Fig. 1. Robotic Laparoscopic grasper model

Grippers are the surgical robots' end-effectors which are in different shapes and have various applications. They are used to grab, cut and stitch. Figure. 2 and 3 show the prevalent surgical grippers and this paper's proposed one respectively.

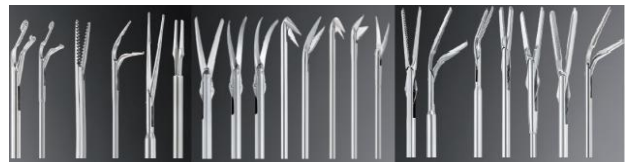

Fig. 2. Various types of conventional grippers used in surgical tools[6]

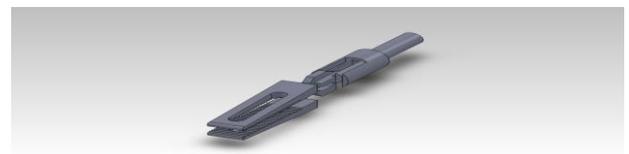

Fig. 3. Model of proposed laparoscopic grasper

Grasper consists of a spring and a towing cable. Spring connects the gripper to connecting rod and makes the movement of gripper possible around the connecting rod. Also 5 towing cables transfer power from servo motors to gripper by passing through the rod so that rotation of the motors to left and right pull cables and this movement leads to different motions in the gripper. The connecting rod is made of stainless 316 steel. One end of the rod is connected to a cylindrical box containing servo motors and the other end is connected to the gripper. The connecting rod is covered by a plastic coating. In the designed laparoscopic grasper, glue is selected to connect different metallic part together, because it is be able to easily disassemble and connect different parts. Obviously this can't be done if we soldier or connect them using screw, because the utilized parts have small dimensions and it is not possible to create screw threads in the inner surface of the rod and external surface of the gripper.

Three gears are used to move the connecting rod where two of them are directly linked to servo-motor and the middle one is connected through a leadscrew.

Whole cylinder and connecting rod would be rotated when upper and lower arms are fixed in their place. Middle servo motor is used to pull and push the cable connected to gripper that makes it to open and close. Two other servo motors are used to pull the cables connected to the spring that provide remaining two degrees of freedom for the grasper robot. The assembled model of gripper is illustrated in Figure.4.

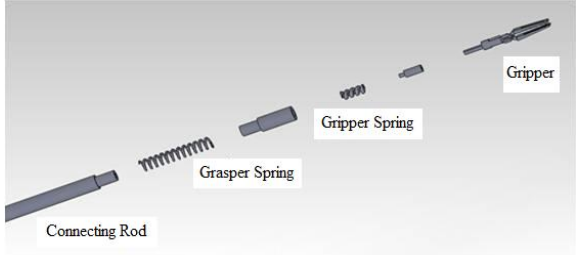

Fig. 4. The assembled model of gripper

## 3. Dynamic model and geometric calculation of grasper robot

Dynamic modeling of surgical robots has two general methods: Energy-based methods and the methods based on Newton formulation. To determine position of the robots' end-effector in Cartesian coordinate relative to ground, the Denavit Hartenberg [16] parameters can be used.

To do this and only to do the calculations, grasper's spring is replaced by three joints. As represented in Figure. 5, two of these joints are considered as revolute to provide rotations around $z_1$ and $z_2$ axes and one of them is considered as prismatic in the direction of $z_3$ to present the elongation of the spring as a result of rotations. Because length of the spring remains as constant, equal to $L_2$, and doesn't experience elongation or compression, this length ($L_2$) would be added to length of the connecting rod ($L_1$) in calculations and thus revolute and prismatic joints can be considered in the distance of $L_2$ from beginning of the spring.

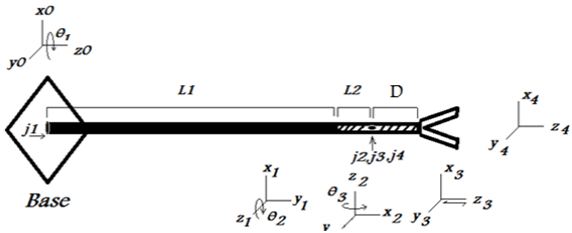

Fig. 5. Geometrical structure of grasper robot

Equation. 1 shows the dynamic model of the gripper robot:

$$M(q)\ddot{q} + C(q,\dot{q})\dot{q} + g(q) = u \qquad (1)$$

where $M(q) \in R^{n \times n}$ is the inertia matrix, $C(q,\dot{q})\dot{q} \in R^n$ is the centripetal and Coriolis matrix, $g(q) \in R^n$ is the gravitational force and u is the exerted joint input. $q \in R^n$ is the joint angle vector. Even if the equations of motion of the robot are complex and highly nonlinear, there are still some basic properties in Eq. (1) that are convenient for controller design. These properties are as follows.

**Property 1.** The inertia matrix M(q) is uniformly positive definite, and there exist positive constants $\alpha_1$, $\alpha_2$ such that:

$$\alpha_1 I \leq M(q) \leq \alpha_2 I \qquad (2)$$

**Property 2.** The manipulator dynamics (1) is linear in a set of physical parameters $\theta_d = (\theta_{d1}, \theta_{d2},..., \theta_{dp})^T$.

$$M(q)\ddot{q} + C(q,\dot{q})\dot{q} + g(q) = Y(q,\dot{q},\ddot{q})\theta_d \qquad (3)$$

Where $Y(q,\dot{q},\ddot{q}) \in R^{n \times p}$ is usually called the dynamic regressor matrix.

The generalized end-effector's position $x \in R^n$ can be expressed as

$$x = h(q) \qquad (4)$$

Where $h(.) \in R^n \to R^n$ is generally a nonlinear mapping between joint space and task space.

Generally, the problem is generating the joint torques $\tau(t)$ such that the robot joint motions q and $\dot{q}$ track the desired trajectories. The goal of the robot controller design is to make the system insensitive to payload and parameter variation as well as to decouple the nonlinear dynamic. In this paper the problem of rejecting a constant bounded tremor is considered. Assume that $d$ is the constant bounded vector describing the surgeons' tremor; the motion equation of the manipulator can be written as:

$$D(q)\ddot{q} + C(q,\dot{q})\dot{q} + G(q) = u + d \qquad (5)$$

$d$ is considered as an unknown parameter and can be shown as:

$$\exists\, 0 \leq \delta < \infty;\ \sup_{0 \leq t < \infty} \|d(t)\|_2 < \delta \qquad (6)$$

In linear systems, one common way for rejecting a bounded disturbance is adding a corresponding internal model of it.

## 4. Manipulator control law formulation

As it was mentioned in pervious section, the robots used in surgery systems interact with patient's body, which is assumed as the environment. Tremor of surgeon or operator's hand can be considered as a disturbance input, whose effect can be eliminated by designing an appropriate controller. In this section it will be tried to find a solution to this problem. The surgeon's hand tremor will be studied as a constant-bounded disturbance. This disturbance will be considered as load torques on input signals. More specifically if $d$ be a constant-bounded vector, then the dynamic equation of motion for the surgical robot can be written as Eq. 1.

### 4.1 Inverse Dynamics Controller with an Integral Action

This method is known as computed torque control in control literature. The basic idea is to seek a nonlinear feedback control law to cancel exactly all of the nonlinear terms in Eq. (5), so that in the ideal case, the closed-loop system is linear and decoupled. The novelty of the paper is rejecting tremor of surgeons that is considered as constant bounded disturbance of system and decoupling links of robot with using inverse dynamic control law with an integral action. For trajectory tracking in task space, resolved acceleration control can be adopted,

$$u = D(q)v + C(q,\dot{q})\dot{q} + G(q) \quad (7)$$

where $v$ is an auxiliary control input that is given by

$$v := \ddot{q}_d(t) + K_D(\dot{q}_d(t) - \dot{q}) + K_p(q_d(t) - q) + K_I \int_0^t (q_d(s) - q)ds \quad (8)$$

Where $K_p, K_D, K_I$ are positive definite diagonal matrices and $q_d(t)$ is desired end-effector trajectory. Assumption $\tilde{q}(t) = q_d(t) - q(t)$ and substituting (7) and (8) into (5) will lead to a linear error dynamic equation

$$\dddot{\tilde{q}} + K_D \ddot{\tilde{q}} + K_p \tilde{q} + K_I \int_0^t \tilde{q}\,ds = \delta \quad (9)$$

where $\delta := D^{-1}(q)d$, $\tilde{q} = q - q_d(t)$. Without loss of generality it is assumed that

$$K_D = k_d I_{n \times n}, \quad K_P = k_p I_{n \times n}, \quad K_I = k_i I_{n \times n} \quad (10)$$

Now for each joint we have

$$\tilde{q}_i = \frac{s}{s^3 + k_d s^2 + k_p s + k_i} \delta_i \ ; \ i = 1,...,n \quad (11)$$

It is easy to show that the final result of the tracking error will be

$$\tilde{q}_i(\infty) = \lim_{s \to 0} \frac{s}{s^3 + k_d s^2 + k_p s + k_i} \delta_i = 0 \quad (12)$$

Since $K_p, K_D, K_I$ are positive definite diagonal matrices thus it is obvious that, the closed-loop system is linear, decoupled and asymptotically stable and inverse dynamics controller with integral action can reject constant bounded tremor successfully.

## 5. Simulation

The control goal is tracking desired trajectory in joint space in the presence of surgeon's hand tremor.
Control law without integral action and disturbance consideration is

$$u = M(q)\left(\ddot{q}_d(t) + K_d(\dot{q}_d(t) - \dot{q}(t)) + K_p(q_d(t) - q(t))\right) + C(q,\dot{q})\dot{q} + G(q)$$

(13)

Proposed gains are determined as $K_p = 4.2 \times I_2, K_d = 2.4 \times I_2$.

Figure. 6 depicted performance of motion trajectory of surgical robot using inverse dynamic method.

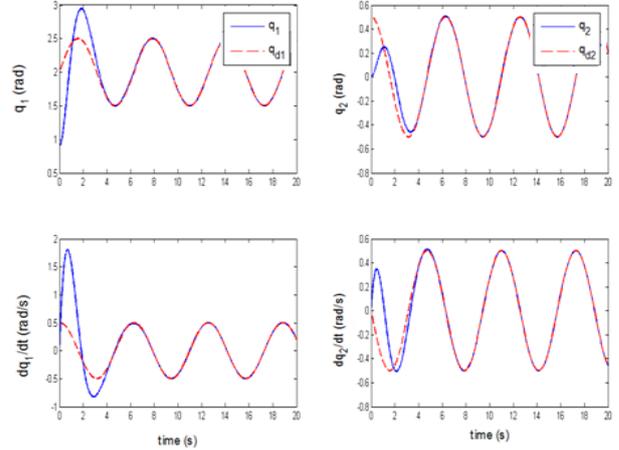

Fig. 6. Trajectory performance based on the inverse dynamics controller without an integral action and disturbances

Tremor modeling leads to terms that are added with the input. In this section inverse dynamic controller will be simulated and their results will be compared in different gains. Inverse dynamic controller with an integral action can be obtained by adding an integrator to the inverse dynamic controller as (7) and (8). Similar to other inverse dynamic controllers, closed loop stability analysis can be done using Horowitz stability criterion. By assuming control gains $K_d = 4.2 \times I_2, K_p = 2.4 \times I_2, K_I = I_2$ and a disturbance $\Delta \tau = \begin{bmatrix} 1 & 0.5 \end{bmatrix}^T$ the following results can be obtained for $q_d(t) = \frac{1}{2}\begin{bmatrix} \sin(t) \\ \cos(t) \end{bmatrix}$, $t \geq 0$.

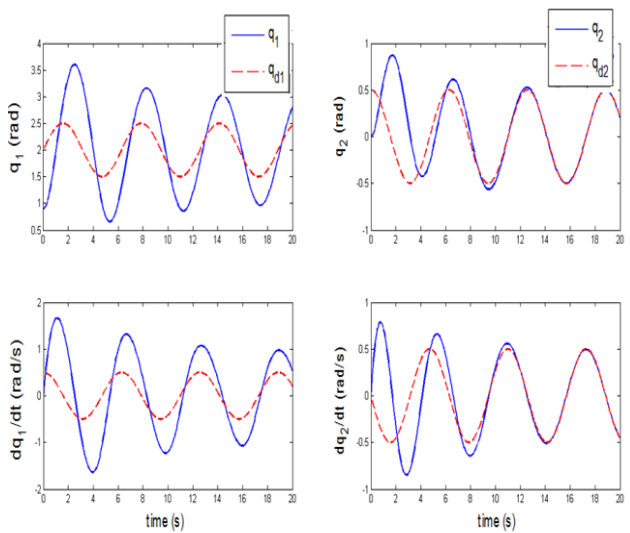

Fig. 7. Trajectory performance based on the inverse dynamics controller with an integral action in the presence of tremor

As it can be seen from Figure. 7, the disturbance rejection process is slow. To overcome this problem and increasing response time, control gains are increased to $K_d = 21 \times I_2, K_p = 12 \times I_2, K_I = 5 \times I_2$.

Figure. 8 presents the tracking performance with above gains. The most important drawback of this robust controller is its sensitivity of gains against magnitude of the disturbance, which means that for a disturbance with larger magnitude, the gains should be increased respectively. The other drawback of this controller is its non-optimality when control gains are increased. Figure 9 represents the control energy in both cases.

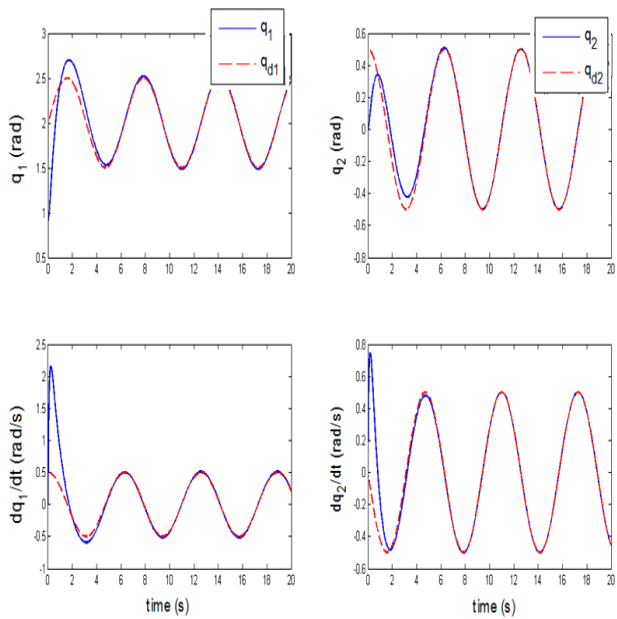

Fig. 8. Trajectory performance based on the inverse dynamics controller with an integral action and high gains in the presence of tremor

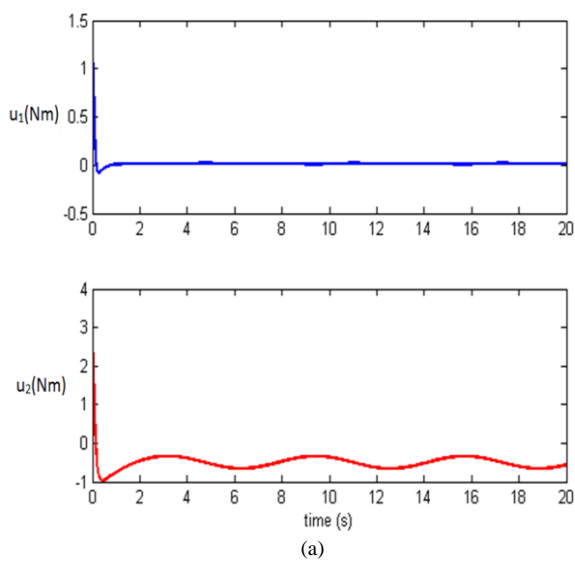

(a)

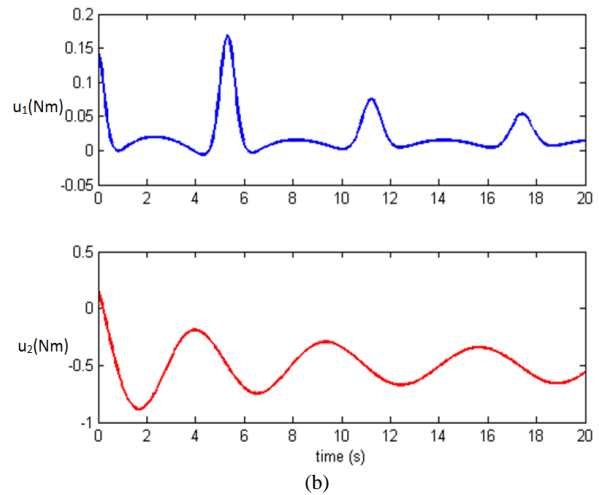

(b)

Fig. 9. Control input signals based on inverse Dynamics controller a) High gain, b) Low gain

## 6. Conclusion

Inverse dynamic controller with an integral action is presented in this paper to improve the accuracy of tracking procedure for surgical manipulators with constant bounded disturbances.

The tracking control problem is formulated as a disturbance rejection problem, with all the system nonlinearities and uncertainties lumped into disturbance.

By comparing inverse dynamics controller with an integral action and Lyapunov –based controller with uncertainty observer, it can be stated that inverse dynamic controller has the capability to be modified, whilst Lyapunov-based controller cannot be easily modified due to its structural complexities. But from another viewpoint, the Lyapunov-based controller is efficiently robust to uncertainties due to its nonlinear nature, where inverse dynamic controller is very sensitive to parameter uncertainties.

The proposed control algorithms were found to generate superior tracking performance and smoother control action.

As a practical example, a grasper of laparoscopic robot is used to assess the performance of proposed methods. Simulation results confirm that control schemes provide an effective means of obtaining high performance trajectory tracking and show good parameter convergence.

## References

[1] R. Muradore, D. Bresolin, L. Geretti, P. Fiorini, T. Villa, "Robotic surgery" *Robotic and Automation Magazine, IEEE*, Volume 18, Issue 3, 2011, Pages 24-32.
[2] C.G.L Cao, E. Danahy, "Increasing accessibility to medical robotics education" *in Proceedings IEEE conf. on Technologies for Practical Robot Applications*, 2011, Pages 49-53.
[3] M. Catenacci, R.L. Flyckt, T. Falcone, "Robotics in reproductive surgery: Strengths and limitations" *Placenta*, Volume 32, Supplement 3, September 2011, Pages S232-S237.

[4] M. Hadavand, A. Mirbagheri, H. Salarieh, F. Farahmand "Design of a force-reflective master robot for haptic telesurgery applications:Robomaster1" *Engineering in Medicine and Biology Society,Annual International Conference of the IEEE*, 2011, Pages 305-310.

[5] Kim, K.-Y.; Song, H.-S.; Suh, J.-W.; Lee, J.-J., "A Novel Surgical Manipulator with Workspace-Conversion Ability for Telesurgety" *Mechatronics, IEEE/ASME Transactions*, Volume 18, Issue 1, 2013, Pages 200-211.

[6] M. Tavakoli, R V. Patel, M.Moallem, A.Aziminejad, *Haptics for Teleoperated Surgical Robotics Systems*. World Scientific Publishing Co. Pte. Ltd, 2008.

[7] G. S. Chirikjian. "Hyper-redundant manipulator dynamics: A continuum approximation". *Advanced Robotics*, Vol. 9, No. 3, pp. 217-243, 1994.

[8] G. Hillerstorm, "Adaptive suppression of vibrations – a repetitive control approach" *IEEE Transactions on Control Systems Technology* 1996, pp. 72-78.

[9] M. Bodson, J.S. Jensen, S.C. Douglas, "Active noise control for periodic disturbances" *IEEE Transactions on Control Systems Technology*, 2001, pp 200-205.

[10] K.B. Ariyur, M. Krstic, "Feedback attenuation and adaptive cancellation of blade vortex interaction on a helicopter blade element" *IEEE Transactions on Control Systems Technology, 1999, PP. 596-605*.

[11] W. M. Lu, "Rejection of persistent $L_\infty$ bounded disturbances for nonlinear systems," *IEEE Trans. Automa. Contr. vol. 43(12), 1998,pp. 1692-1702*.

[12] M. Wilson, "The role of seam tracking in robotic welding and bonding", The Industrial Robot. vol. 29, no. 2, pp.132-137, 2002.

[13] P. E. Wellstead ,M. B. Zarrop, Self-Tuning Systems: Control and Signal Processing, John Wiley a Sons,1991.

[14] Colbalugh, R., Glass, K., Seraji, H., "Performance-Based Adaptive Tracking Control of Robot Manipulators," J.Robotic System, vol. 12, pp. 517-530, 1995.

[15] Reed, J., and Ioannou, P., "Instability Analysis and Robust Adaptive Control of Robotic Manipulator," IEEE Trans. Robotics and Automation, vol. *5,* no. 3,, pp. 381-386 ,1989.

[16] Mark W.Spong, Seth Hutchinson, M Vidyasagar, Robot Modeling and Control. Wiley Press, 2006.

[17] H. Khalil, Nonlinear systems, Third edition, Prentice hall, 2001.